\documentclass[11pt,twoside]{article}
\usepackage{asp2010}

\resetcounters

\begin{document}

\title{The Effective temperature scale of M dwarfs from spectral synthesis}
\author{C\'eline Reyl\'e$^1$, Arvind S. Rajpurohit$^1$,  Mathias Schultheis$^1$, and France Allard$^2$
\affil{$^1$Observatoire de Besan\c{c}on, Institut Utinam, UMR 6213 CNRS, BP 1615, F-25010 Besan\c{c}on Cedex, France}
\affil{$^2$Centre de Recherche Astrophysique de Lyon, UMR 5574 CNRS, Universit\'e de Lyon, \'Ecole Normale Sup\'erieure de Lyon , 46 All\'ee d'Italie , F-69364 Lyon Cedex 07, France}}

\begin{abstract}
We present a comparison of low-resolution spectra of 60 stars covering  the whole M-dwarf sequence. Using the most recent PHOENIX BT-Settl stellar model atmospheres (see paper by F. Allard, in this book) we do a  first quantitative comparison to our observed spectra in the wavelength range 550-950 nm.  We perform a first confrontation between models and observations and we assign an effective temperatures to the observed M-dwarfs. Teff-spectral type relations are then compared with the published ones. This comparison also aims at improving the models' opacities.
\end{abstract}

\section{Introduction}
Low-mass dwarfs are the dominant stellar component of the Galaxy. Our understanding of the Galaxy therefore relies upon the description of this faint component. Indeed M-dwarfs have been employed in several Galactic studies. Moreover, M dwarfs are now known to host exoplanets, including super-Earth exoplanets \citep{Bonfils2007,Udry2007}.
The determination of accurate fundamental parameters for M dwarfs has therefore relevant implications for both stellar and Galactic astronomy. 

Over the last decade, stellar models of very low mass stars have made great progresses. One of the most important recent improvements is the availability of new atomic line profile data that give a much improved representation of the details of the line shapes in the optical spectra of cool dwarfs, and become especially important in situations where line blanketing and broadening are crucial for the model.
Still models have to use some incomplete or approximate input physics such as uncertain oscillator strengths for some line and molecular bands missing opacities sources (VO, FeH, CaOH, and some TiO bands). Descriptions of these stars therefore need a strong empirical basis, or validation.

In section~\ref{obs}, we present a comparison of low-resolution spectra covering the whole M-dwarf sequence with recent stellar atmosphere models. In section~\ref{teff}, we derive a spectral type - effective temperature relation and compare it with the published ones.

\section{Comparison between atmosphere models and M-dwarf spectra}
\label{obs}

M-dwarfs remain elusive and enigmatic objects because of their small size and cool surface temperature. M-dwarf spectra are characterized by the presence of strong molecular absorptions such as TiO, VO, H$_2$O and CaH. The temperatures, abundances, sizes and luminosities are not yet well understood. We compared 60 M-dwarfs (from M0 to M9) with optical spectroscopic classification \citep{Reyle2006}, on a large wavelength range, with the most recent PHOENIX stellar atmosphere models, varying the effective temperature, gravity and metallicity. The models used are the most recent version BT-Settl 2010 (Allard, Homeier \& Freytag, in this book),   taking into account : i) the solar abundances revised by \cite{Asplund2009}, ii) the most recent BT2 version of the water vapor line lists by \cite{Barber2008}, iii) a cloud model based on condensation and sedimentation timescales by \cite{Rossow1978}, supersaturation computed from pre-tabulated chemical equilibrium, and mixing from 2D radiation hydrodynamic simulations by \cite{Freytag2010}. The models are available on-line.\footnote{http://phoenix.ens-lyon.fr/simulator}.

\begin{figure}[ht]
\begin{center}
\plotone[scale=0.6,angle=0,clip=]{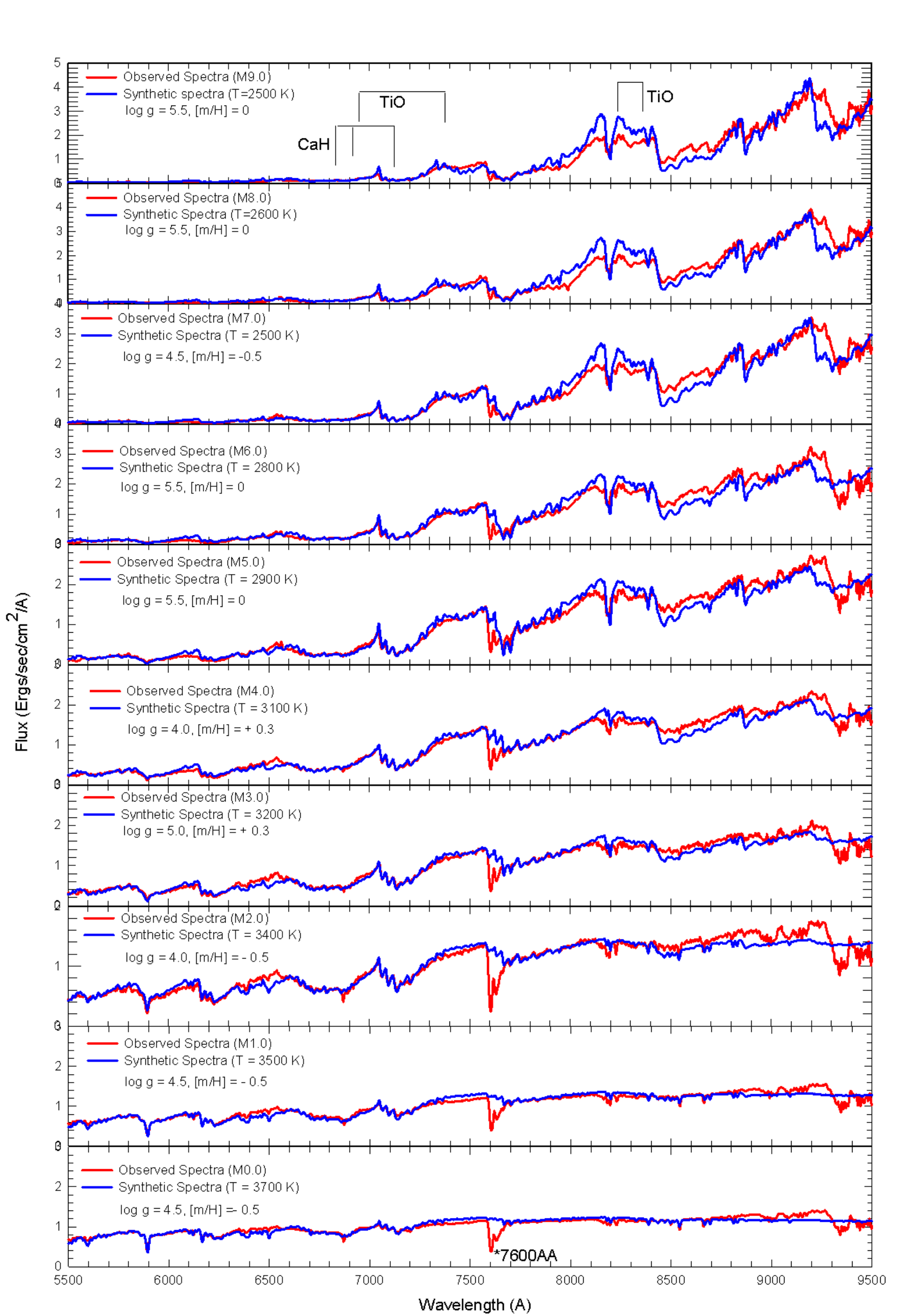}
\caption{Comparison between model (blue line) and observations (red line) for M0 (bottom) to M9 (top) dwarfs. The strong telluric line around 7600\AA\ is also indicated.}
\label{fig1}
\end{center}
\end{figure}

Fig.~\ref{fig1} shows the comparison between model (blue line) and observations (red line) along the M-dwarf sequence, from M9 (top) to M0 (bottom). 
The slope of the optical to near IR spectra is well reproduced by the BT-Settl 2010 models, while some discrepancies remain in the strength of some absorption bands such as the TiO absorption around 6500\AA. The quality of the fit deteriorates as one goes from the early M to the late Ms and early L dwarfs. These results will allow to calibrate the missing oscillator strengths of molecular bands.

\section{Effective temperature scale of M-dwarfs}
\label{teff}

\begin{figure}[ht]
\begin{center}
\plotone[scale=0.4,angle=0,clip=]{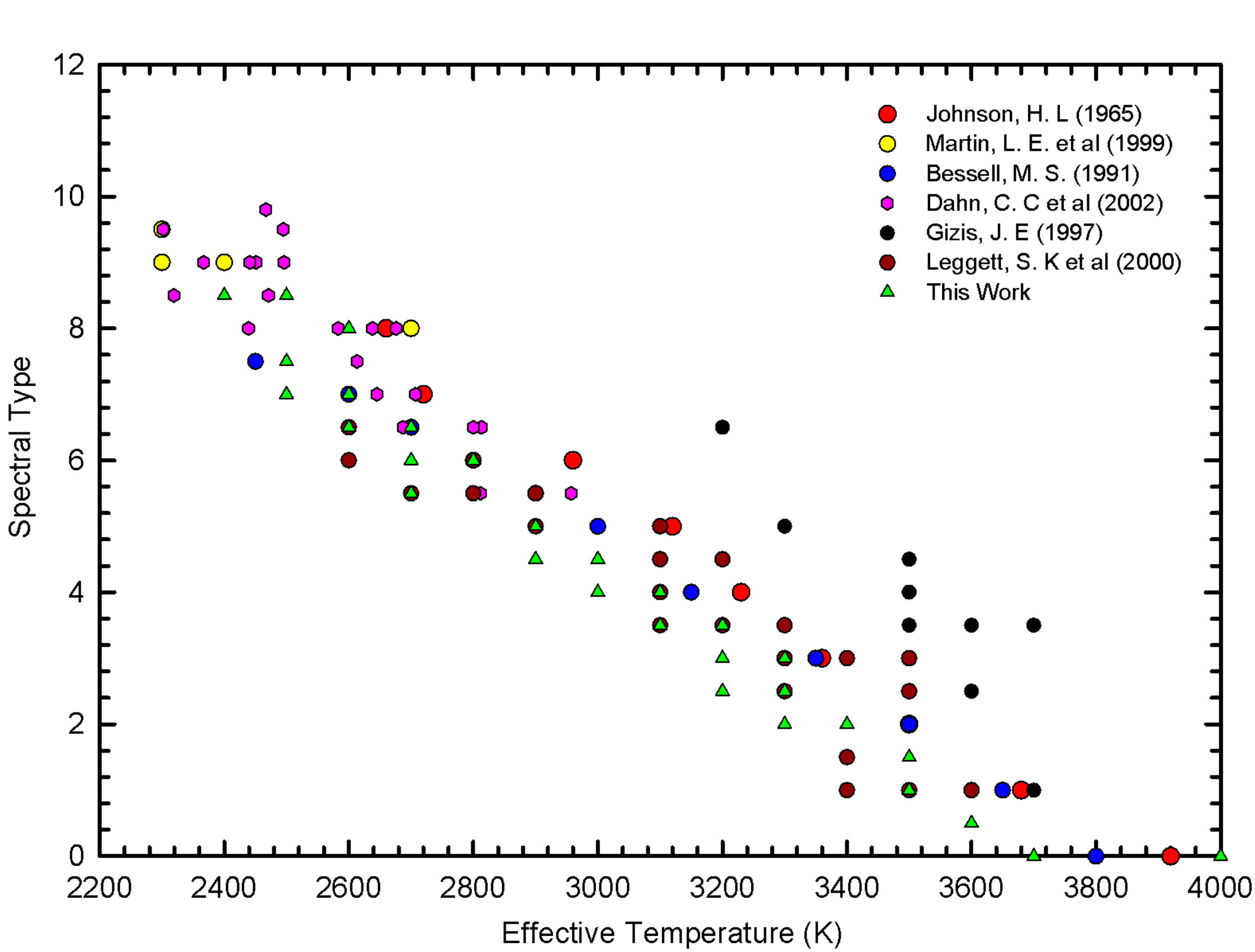}
\caption{Our spectral type versus effective temperature (green triangles) compared to relations from \cite{Johnson1964,Bessel1991,Gizis1997,Martin1999,Leggett2000,Dahn2002}}
\label{fig2}
\end{center}
\end{figure}

We used the grid of models and a $\chi^2$ fitting method to derive the stellar parameters (effective temperature, metallicity, gravity) of our sample. The spectral type are known from optical spectroscopy classification \citep{Reyle2006}, using the spectral indices TiO$_5$, CaH$_2$, and CaH$_3$. Multicolour photometry 
is an alternative way to determinr the effective temperature \citep[see e...g.][]{Johnson1965} but it requires an assumption on the diameter of the star.
Fig.~\ref{fig2} shows the relation between spectral type and effective temperature, compared to others found in the literature.

\section{Conclusion}

We compared 60 spectra in the wavelength range 550-950 nm of M dwarfs with synthetic spectra obtained with the most recent PHOENIX BT-Settl stellar model atmospheres (see Allard et al., in this book). Our sample covers the M-dwarf spectral sequence. We found that the slope of the optical to near IR spectra is well reproduced by the models, while some discrepancies remain in the strength of some absorption bands. The quality of the fit deteriorates as one goes from the early M to the late Ms and early L dwarfs. This comparison will allow to calibrate the missing oscillator strengths of molecular bands. We derived the effective temperature of the 60 M-dwarfs and drew a relation between effective temperature and spectral type, that we compared to others found in the literature.

\acknowledgements Financial support from the ÓFrench Programme National de Physique StellaireÓ (PNPS) is gratefully acknowledged.

\bibliography{cs16reyle}

\end{document}